# Nonlocal nonlinear phononics

M. Henstridge[1], M. Först[1], E. Rowe[1], M. Fechner[1], A. Cavalleri[1,2]

[1] *Max Planck Institute for the Structure and Dynamics of Matter, 22761 Hamburg, Germany*

[2] *Department of Physics, Clarendon Laboratory, University of Oxford, Oxford OX1 3PU, United Kingdom*

**Nonlinear phononics[1] relies on the resonant optical excitation of infrared-active lattice vibrations to coherently induce targeted structural deformations in solids. This form of dynamical crystal-structure design has been applied to control the functional properties of many interesting systems, including magneto-resistive manganites[2], magnetic materials,[3-5] superconductors[6,7], and ferroelectrics[8-10]. However, phononics has so far been restricted to protocols in which structural deformations occur locally within the optically excited volume, sometimes resulting in unwanted heating. Here, we extend nonlinear phononics to propagating polaritons, effectively separating in space the optical drive from the functional response. Mid-infrared optical pulses are used to resonantly drive an 18 THz phonon at the surface of ferroelectric $LiNbO_3$. A time resolved stimulated Raman scattering probe reveals that the ferroelectric polarization is reduced over the entire 50 μm depth of the sample, far beyond the ~μm depth of the evanescent phonon field. We attribute the bulk response of the ferroelectric polarization to the excitation of a propagating 2.5 THz soft-mode phonon-polariton. For the highest excitation amplitudes, we reach a regime in which the polarization is reversed. In this this non-perturbative regime, we expect that the polariton model evolves into that of a solitonic domain wall that propagates from the surface into the materials at near the speed of light.**

Generally, optical radiation tuned to frequencies immediately above the TO frequency of an infrared-active vibration does not propagate into a material and is evanescently screened by the induced polarization. The interaction of the light field with the material, and the ability to manipulate its microscopic properties remains therefore inherently local. Here, we show that if the material breaks inversion symmetry, the optically excited mode can couple anharmonically to other symmetry-odd, infrared active excitations of the solid, which hybridize with the electromagnetic field and form a propagating polariton. This type of nonlinear phonon-polariton coupling between an evanescent drive and a propagating mode opens up a broad class of phenomena that we term *nonlocal nonlinear phononics*.

Nonlocal polariton excitation builds on the extensive work performed in the past in which polaritons were excited using impulsive stimulated Raman scattering from propagating near-infrared pulses[11,12] In these studies, steering, focusing, and amplification of THz radiation was shown to occur in regions of the crystal beyond those traversed by the excitation pulse[13,14]. However, even in cases in which phase matching was optimized to drive phonon polaritons of the largest amplitudes, these experiments never reached the deep nonlinear regime that is afforded by nonlinear phononics. In the present paper we show that nonlinear phononics can bring this class of phenomena to the more consequential regime of ferroelectric switching, predicted[15] but never observed experimentally through polaritons.

The principle of nonlocal nonlinear phononics is sketched in Fig. 1a for the model ferroelectric $LiNbO_3$, which is studied in this work. Both infrared active mode $Q_{IR}$ and polariton $Q_P$ are connected to lattice vibrations. The c-axis polarized 18.3 THz $A_1$ mode ($Q_{IR}$) is nonlinearly coupled to the 7.3 THz $A_1$ "soft mode" ($Q_P$), which in turn modulates the ferroelectric polarization. Previous work has shown that driving the high-frequency vibration $Q_{IR}$ with a resonant sub-picosecond mid-infrared pulse results in a transient reversal of the ferroelectric polarization. However, the reversal was only detected below the surface of the material by time resolved second-harmonic generation[16]. The spatial extent of this effect was not accessible by that method, as it was limited by the micrometer coherence length of the detection process. Here, we demonstrate that the ferroelectric polarization in LiNbO$_3$ can be controlled non-locally throughout the bulk. Resonant excitation of Q$_{IR}$ at 18.3 THz also launches an anharmonically coupled low-frequency polariton of the soft mode Q$_P$, which, contrary to the evanescent driven mode, propagates throughout the crystal (Figure 1b). Femtosecond stimulated Raman scattering measurements, which employ short pulses of visible light below the optical bandgap, interrogate the lattice of LiNbO$_3$ and reveal a reduction of the Raman cross section for a number of polar modes, which disappear transiently as the ferroelectric polarization vanishes. At the highest fields, we even observe that the Raman susceptibility of these modes changes sign, indicating a reversal of the ferroelectric polarization throughout the bulk of the solid, and in fact far beyond the evanescent field of the mid-infrared optical drive.

To describe these physics quantitatively, we consider the following potential for the dynamics of the crystal lattice:

$$V(Q_{IR}, Q_P) = \frac{1}{2}\omega_{IR}^2 Q_{IR}^2 + \frac{1}{2}\omega_P^2 Q_P^2 + V_{AH}(Q_{IR}) + V_{AH}(Q_P) + g Q_{IR}^2 Q_P \quad (1)$$

where $V_{AH}(Q_{IR,P}) = \frac{1}{3}a_{IR,P} Q_{IR,P}^3 + \frac{1}{4}b_{IR,P} Q_{IR,P}^4$ are high-order nonlinearities along the phonon coordinates $Q_{IR}$ and $Q_P$, relevant during high amplitude excitation[17]. The last term describes the lowest-order anharmonic coupling between $Q_{IR}$ and $Q_P$ which enables energy transfer between the two modes. The polarization in the crystal is given by:

$$P = \epsilon_o \epsilon_\infty E + P_{ion} + P_{Ram} \quad (2)$$

where $\epsilon_o$ is the vacuum permittivity and $\epsilon_\infty$ is the high frequency dielectric constant. $P_{ion} = Z_{IR}^* Q_{IR} + Z_P^* Q_P$ is the ionic contribution to the polarization resulting from the effective charges $Z_{IR}^*$ and $Z_P^*$ of the modes $Q_{IR}$ and $Q_P$, respectively. $P_{Ram} = R_{IR}^* Q_{IR} E + R_P^* Q_P E$ is the polarization which results from the coupling of $Q_{IR}$ and $Q_P$ to the total electric field E via the respective Raman tensors $R_{IR}^*$ and $R_P^*$. Assuming a light-matter interaction given by $\boldsymbol{P} \cdot \boldsymbol{E}$, the respective equations of motion are:

$$\ddot{Q}_{IR} + 2\gamma_{IR}\dot{Q}_{IR} + \omega_{IR}^2 Q_{IR} + a_{IR} Q_{IR}^2 + b_{IR} Q_{IR}^3 = Z_{IR}^* E + R_{IR}^* E^2 - 2g Q_{IR} Q_P \quad (3)$$

$$\ddot{Q}_P + 2\gamma_P \dot{Q}_P + \omega_P^2 Q_P + a_P Q_P^2 + b_P Q_P^3 = Z_P^* E + R_P^* E^2 - g Q_{IR}^2 \quad (4)$$

An incident electric field thus imparts two forces upon $Q_{IR}$ and $Q_P$: one that can couple directly to the modes through their effective charges *and* a "Raman" force, which we found to be much weaker by comparison. Notably, the rectified component of the $gQ_{IR}^2$ term in eqn. (4) imparts a quasi-static force onto $Q_P$ when $Q_{IR}$ is driven to high amplitudes.

Using finite difference time domain methods, the nonlinear lattice response of LiNbO$_3$ to an incident mid-infrared pulse of duration 200 fs and central frequency 20 THz is readily simulated. Figure 1(c) shows the time and space dependent mode amplitudes $Q_P$ and $Q_{IR}$ in the regime of low-field (0.1MV/cm) excitation. The frequency components of $Q_{IR}$ excited at $\omega_{IR}$ and within the Reststrahlen band have the largest amplitudes and penetrate the material evanescently to depths on the order of a few microns. The frequency components of $Q_{IR}$ driven by the broadband excitation pulse immediately below $\omega_{IR}$ correspond to a polariton which propagates throughout the material. This polariton radiates an electric field which applies a linear force onto $Q_P$, causing it to also oscillate across the bulk. Importantly, all features shown in Figure 1(c) are the result of a linear excitation in which anharmonic phonon coupling is negligible; there are no frequency components outside the bandwidth of the excitation pulse.

Figure 1(d) shows $Q_{IR}$ and $Q_P$ in the case of a high field (30 MV/cm) excitation. The features from the linear excitation regime are still present, however, the nonlinear force on $Q_P$ arising from the $gQ_{IR}^2$ term in eqn (4) is now considerable. The rectified component of this force excites a low-frequency (2.5

THz) mode of the $Q_P$ polariton branch shown in Figure 1(b), resulting in a distinct polarization wave propagating throughout the material. As discussed below, we assert that the control of the ferroelectric polarization in the bulk of LiNbO$_3$ is intertwined with the propagation of this ferroelectric soft mode polariton.

In our experiments, we studied the response of the Raman-active lattice modes throughout a 50 micron thick LiNbO$_3$ single crystal using femtosecond stimulated Raman spectroscopy (FSRS)[18-20]. The theory behind FSRS has been described elsewhere[21,22]. Here, we briefly summarize the general concept.

Figure 2 shows a schematic of an FSRS probe experiment and an example of the resulting signal. FSRS requires two incident electric fields, $E_r$ and $E_w$, with respective carrier frequencies $\omega_r$ and $\omega_w$ typically in the visible or near-infrared spectral range. $E_r$ has a narrow spectral bandwidth (< 1 THz) corresponding to a Gaussian envelope in time with full width at half-maximum (FWHM) of typically 1-3 ps. $E_w$ is typically a coherent white light continuum pulse with a spectral bandwidth spanning tens of THz and temporal FWHM ≲ 100 fs. When the two pulses impinge upon a given material, a force $F = R^* E^2$ is then imparted onto a given Raman-active mode $Q$ with eigenfrequency $\omega_Q$. In FSRS, we are concerned with the force component proportional to $E_r E_w$ which oscillates at the difference frequency $\omega_r - \omega_w$. Any mode with frequency $\omega_Q$ can be stimulated as long as $E_w$ contains spectral weight at $\omega_r \pm \omega_Q$. The stimulated mode $Q$ then mixes with $E_r$ via the Raman tensor, generating a nonlinear polarization $P_{rQ} = R^* E_r Q$ which oscillates at frequencies $\omega_r \pm \omega_Q$ and propagates in the same direction as $E_w$. When sent to a spectrometer, the field emitted from the polarization interferes with $E_w$, resulting in amplification at frequency $\omega_r - \omega_Q$ and depletion at $\omega_r + \omega_Q$. In the limit of a static Raman tensor, FSRS cannot discern whether a Raman tensor element is positive or negative, as in the case of an incoherent Raman scattering probe. However, contrary to the case of incoherent scattering, FSRS can detect a transient sign reversal of the Raman susceptibility. If $R^*$ experiences a sign change between the excitation and mixing processes, then the detected Raman peaks reverse sign.

Thus, this particular form of coherent Raman scattering is well-suited for measuring the dynamics of the Raman tensor elements, which, as discussed later, in the particular case of LiNbO$_3$ are correlated to the magnitude and sign of the ferroelectric polarization.

A 50 micron-thick x-cut LiNbO$_3$ single crystal was illuminated with mid-infrared pulses of central frequency of 20 THz, Gaussian intensity FWHM of 175 fs, and repetition rate of 1 kHz. The impinging field was polarized along the c-axis in order to excite the highest frequency A$_1$ mode at 18.3 THz. The FSRS probe was designated to measure frequency shifts relative to a wavelength of 400 nm ($\omega_r$ = 750 THz). To probe the response of the A$_1$ modes, both $E_r$ and $E_w$ were polarized along the c-axis, and the time delay between these two pulses was kept fixed. An FSRS spectrum was recorded for each pump-probe delay $\tau$ between the FSRS pulse pair and the mid-infrared pulse as depicted in Figure 3(a). Further details on the experimental setup are provided in the supplementary information[23].

Figure 3(b) shows the evolution of the FSRS spectra with the time delay $\tau$ upon illumination with a mid-infrared pulse of fluence 38 mJ/cm$^2$. The two most notable changes to the spectra are highlighted in Figure 3(c), which displays the pumped spectrum for $\tau = 0.05$ ps together with the equilibrium response. The amplitudes of the equilibrium Raman peaks reduce (highlighted in red) and several new peaks appear (indicated in blue). Both features can be explained under the assumption that the Raman tensors $R_{IR}^*$ and $R_P^*$ change dynamically across the bulk of the sample upon mid-infrared excitation.

For comparison, Figure 4(a) shows calculated FSRS spectra. The new Raman peaks in blue were produced by imparting a transient 16 THz modulation upon both $R_{IR}^*$ and $R_P^*$, propagating at the speed of the Q$_{IR}$ polariton into the sample. When probed with FSRS, the Raman mode amplitudes $Q_{IR}$ and $Q_P$ then become frequency-modulated at 16 THz, resulting in peaks at $\omega_{IR} \pm 16$ THz and $\omega_P \pm 16$ THz. This is effectively a form of second-order Raman scattering which results from the excitation of atomic motions along the $Q_{IR}$ coordinates just below the TO frequency $\omega_{IR}$. The group velocity at the 400 nm wavelength of the FSRS probe matches the phase velocity of the $Q_{IR}$ polariton at 16 THz[23], thus allowing for the effective detection of effects on the Raman tensor at this modulation frequency.

More important is the reduction in the peak amplitudes of $Q_{IR}$ and $Q_P$. This effect was well-modelled under the assumption that the equilibrium components of both $R_{IR}^*$ and $R_P^*$ become transiently reduced upon excitation with the mid-IR pulse. As shown in Figure 4(b), the group velocity of the probe around 400 nm wavelengths approximately matches the soft mode polariton phase velocity over the course of the 50 µm sample volume. Thus, reductions in the Raman tensor correlated to the propagating soft mode polariton can be detected.

To produce the calculated pumped spectra shown in Figure 4(a), we assume that the Raman tensors become reduced over a time scale similar to those of the transient polarization reductions observed in Ref. 8. Our model then assumes that these reductions of the Raman tensor propagate throughout the crystal at the phase velocity of the soft mode polariton. To account for the propagation mismatch with the probe, the velocity line from Figure 4(b) was projected onto the space-time map for the Raman tensor, and an integral along the velocity line was carried out for each time step. This produces an effective time-dependent Raman tensor which is seen in the rest frame of the probe; the result is highlighted in red in Figure 4(c). A damped sinusoid with a frequency of 16 THz was then added to the result to account for the phased-matched detection of the 16 THz modulation of the Raman tensor. The final curve, shown in black in Figure 4(c), models the effective $R^*(t)$ seen by the probe and was used to produce the spectra shown in Figure 4(a). Complete details regarding the calculations are provided in the supplementary information[23].

The fluence dependence of the peak reductions are shown in Figure 5(a) for a fixed time delay $\tau = 0.08$ ps. As the mid-infrared excitation influence is increased, the FSRS peaks decrease monotonically in amplitude and then reverse sign at the maximum fluence of 185 mJ/cm$^2$ used in the experiment. First principles calculations presented in Figure 5(b) show that *if* the ferroelectric polarization is reversed, the signs of both Raman tensors $R_{IR}^*$ and $R_P^*$ would also reverse. The amplitudes of the FSRS peaks corresponding to $Q_{IR}$ and $Q_P$ as a function of mid-IR pump fluence are shown in Figure 5(c). For fluences above ~ 60 mJ/cm$^2$, the peak integrals scale nonlinearly with fluence and display a saturation-like behaviour before they reverse sign. This is likely due to a softening of the frequency of the driven mode $Q_{IR}$, an effect which occurs when a vibration is driven to high amplitudes and effectively reduces its response to the driving field[17]. The evolution of the peak integrals with time delay τ at maximum excitation fluence are presented in Figures 5(d) and (e) along with calculated values. The calculations were performed in the same manner as those shown in Figure 4, with the exception that here, the Raman tensor is assumed to exhibit a transient sign change. Most importantly, the Raman tensors of both modes reverse sign and then recover to their equilibrium values, evidencing a transient, reversal of the ferroelectric polarization in the bulk crystal.

It should be stated explicitly here that the manipulation of the ferroelectric polarization in the bulk is not described comprehensively by the theory of propagating polaritons which is illustrated in Figure 1. When a ferroelectric domain is either reduced or reversed at the surface, a solitonic phase front of altered domains will also presumably propagate into the bulk at near-luminal speeds. Such a phenomenon breaches the highly interesting regime of driven phase transitions that should be explored further, both theoretically and experimentally.

In summary, by measuring the dynamical Raman tensor elements in LiNbO$_3$ using a stimulated Raman scattering probe, we show control of the ferroelectric polarization well-beyond the excitation region. We extend the ideas of nonlinear phononics to propagating polaritons, which provides a foundation for explaining this phenomenon. Our experiments also provide evidence for a transient reversal of the ferroelectric polarization throughout the bulk of the material, a striking observation which requires further theoretical pursuit.

We foresee that further steps forward could involve integrating nonlinear phononics with nonlocal optical platforms such as metamaterials. This would introduce the ability to refract[24,25], tightly focus[26], and even accelerate[27,28] nonlinearly driven modes. Furthermore, nonlocal nonlinear phononics broadens the scope of optical control in quantum materials and is naturally applicable to other types of propagating polaritons[29]. Finally, we note that the physics discussed here is likely connected to propagating phase changes induced by phononics across hetero-interfaces,[30-32] which to date have remained unexplained.

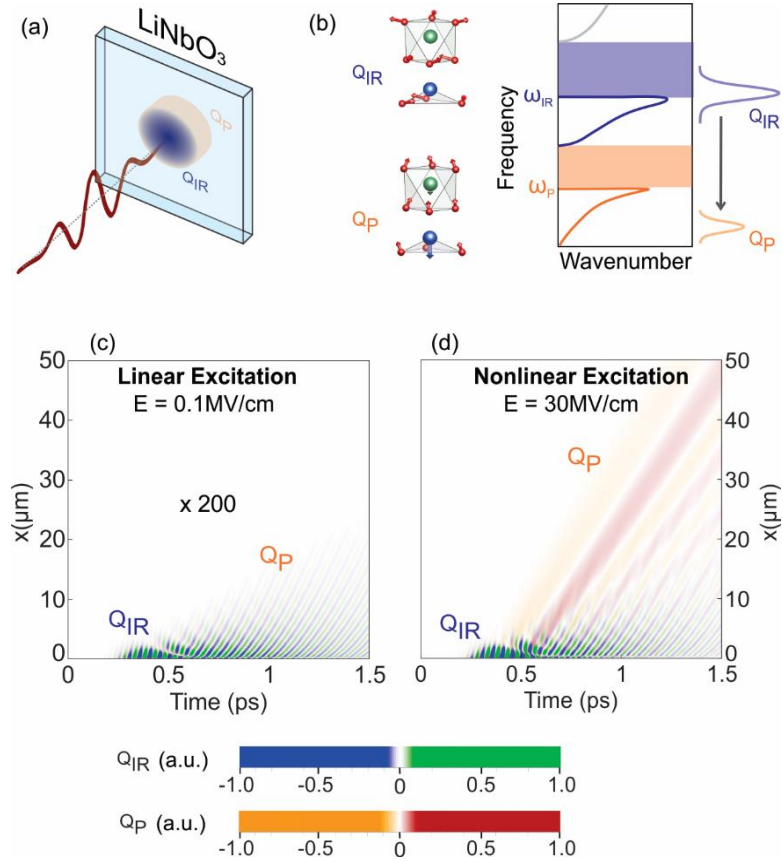

**Fig. 1| Nonlocal nonlinear phononics** (a) An evanescent excitation pulse drives a mode $Q_{IR}$ which launches a propagating mode $Q_P$ via anharmononic phonon coupling. (b) Dispersion relation for LiNbO$_3$ along the crystallographic c-axis. The parameters used to generate the curves were taken from Ref. 33. (c) Simulations of the mode amplitudes $Q_{IR}$ and $Q_P$ upon excitation with a 200fs pulse tuned slightly above $\omega_{IR}$ at 20 THz for an incident field strength of E = 0.1 MV/cm. In addition to the evanescent excitation of $Q_{IR}$ at frequencies within the Reststrahlen band, the frequency components of the pulse which lie just below $\omega_{IR}$ also excite a polariton of $Q_{IR}$ which extends further into the material. However, anharmonic phonon coupling is negligible and no frequency components outside the bandwidth of the incident pulse are produced. (d) Same as (c), except with an incident field strength of 30 MV/cm. Here, the nonlinear force on $Q_P$ becomes appreciable, resulting in the launching of a low frequency polarization wave which propagates throughout the medium.

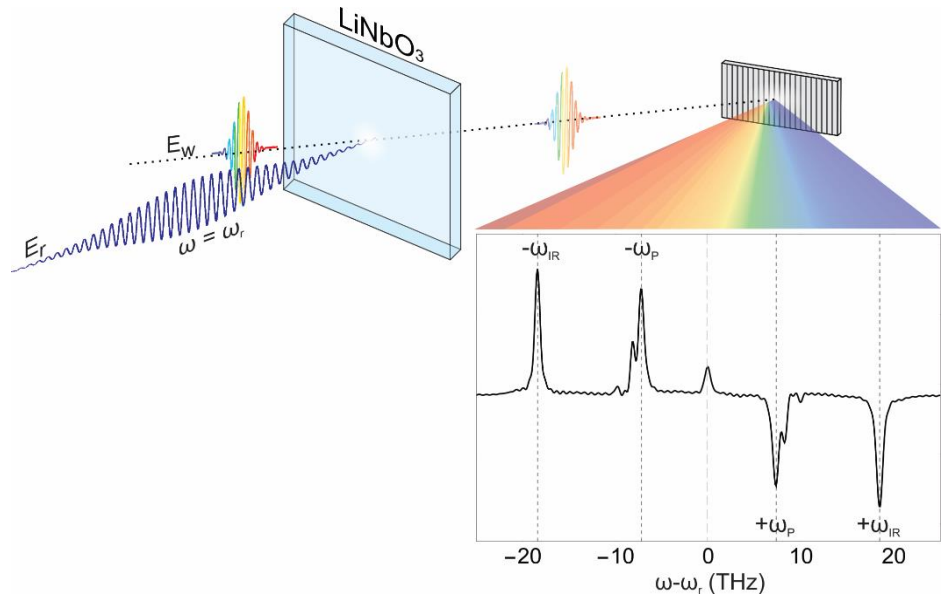

**Fig. 2 | Femtosecond Stimulated Raman Scattering.** A narrowband pulse $E_r$ and a white light continuum field $E_w$ excite a coherent vibration with frequency $\omega_Q$ with a force proportional to $E_r E_w$. The stimulated mode mixes with $E_r$ via the Raman tensor, resulting in the emission of a field with frequencies $\omega_r \pm \omega_Q$ which propagates in the direction of $E_w$. When sent to a spectrometer, the two fields interfere, resulting in amplification at frequency $\omega_r - \omega_Q$ and depletion at $\omega_r + \omega_Q$. Shown is a typical FSRS spectra of the $A_1$ modes in LiNbO$_3$.

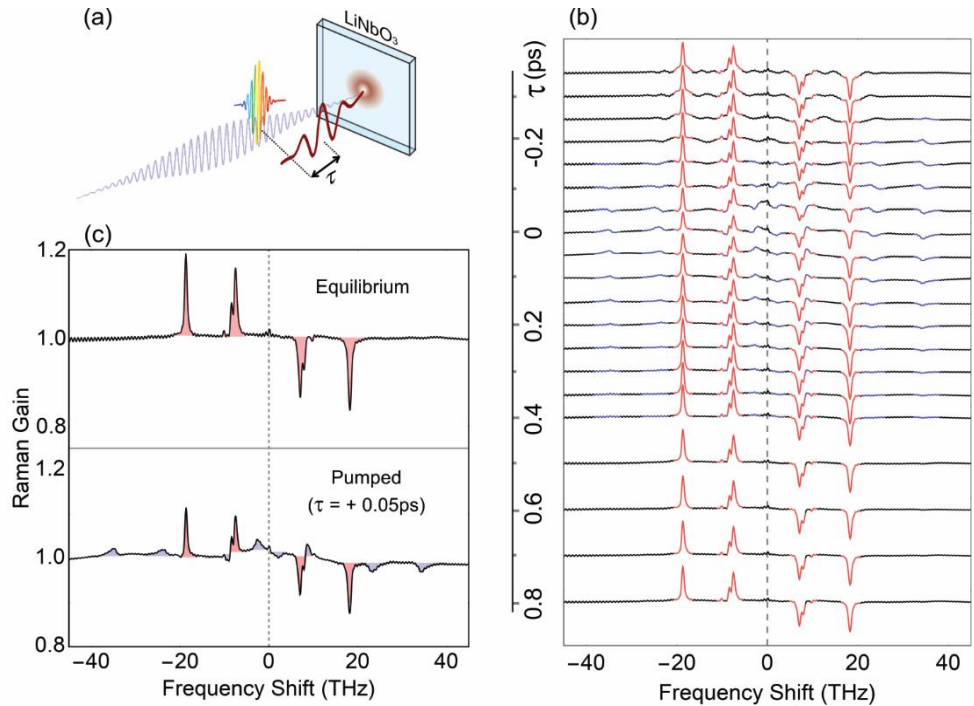

**Fig. 3 | Time-resolved FSRS measurements of the $A_1$ modes in LiNbO3.** (a) Experimental geometry. Time-resolved changes to the FSRS spectra were recorded by varying the time-delay $\tau$ between the FSRS pulse pair and the mid-IR pump pulse. (b) FSRS spectra as a function of $\tau$ for a mid-IR pulse fluence of 38 mJ/cm$^2$. The two prominent features which result from mid-IR excitation are a reduction in the amplitudes of the $A_1$ modes (highlighted in red) and the emergence of new peaks (highlighted in blue). (c) Equilibrium FSRS spectrum and pumped spectrum from (b) for $\tau = 0.05$ ps. As discussed in the text, the features in the pumped spectrum can be explained under the assumption that the Raman tensors for the $A_1$ modes change dynamically.

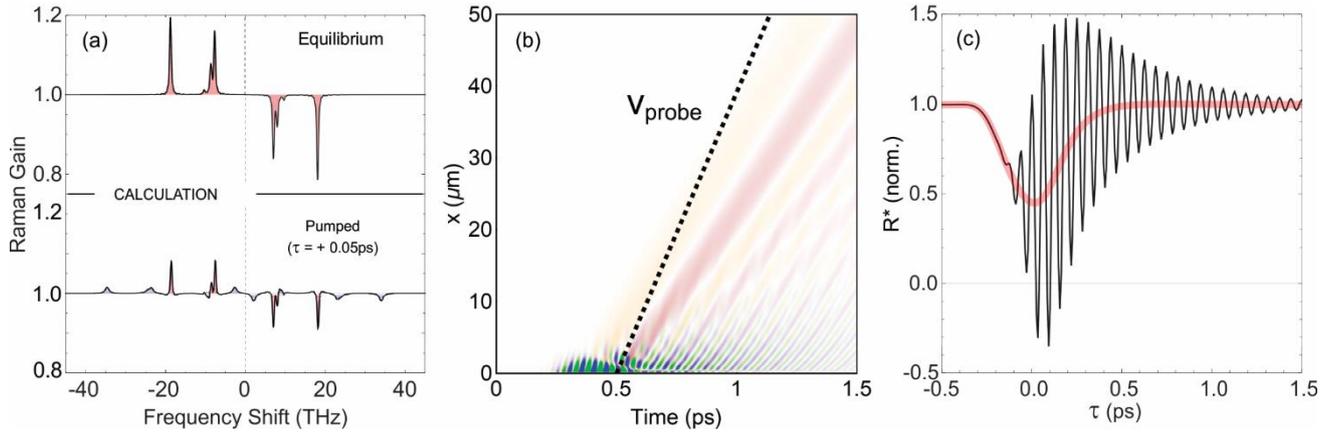

**Fig. 4| Calculations of pumped FSRS spectra in LiNbO$_3$.** (a) Equilibrium and pumped FSRS spectra for τ = 0.05 ps. The calculations were produced using a time-domain model for the FSRS process which is detailed in the supplementary information. (b) Group velocity of the FSRS probe projected onto the space-time maps of $Q_{IR}$ and $Q_P$ from Figure 1(d). Over the course of the 50 μm sample volume, the probe velocity is approximately matched with the phase velocity of the soft mode polariton. Here, the effective velocity of the probe was calculated given the group refractive index of LiNbO$_3$ at 400 nm, the angle between the mid-IR pump and white light continuum, and a correction to the static refractive index of LiNbO$_3$ taken from Ref. 33. (c) Shown in black is the dynamic Raman tensor used to produce the pumped calculations displayed in (a). The red curve indicates the transient reduction of the Raman tensor which is seen in the rest frame of the probe after accounting for the effects of velocity mismatch throughout the 50 μm crystal thickness. The component which oscillates at 16 THz imparts sidebands upon the frequencies of the equilibrium modes, giving rise to the blue-highlighted peaks in (a).

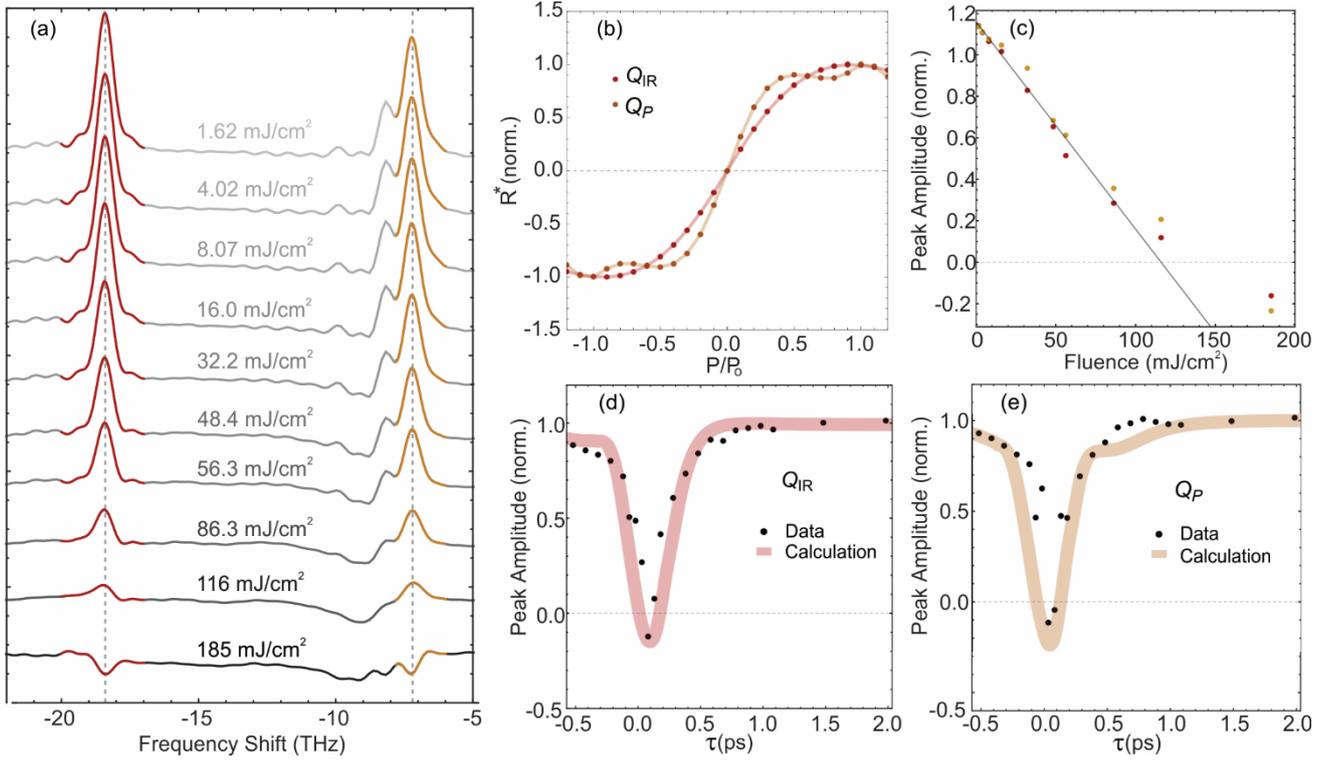

**Fig. 5 | Reversal of the ferroelectric polarization in the bulk.** (a) Fluence dependence of the $A_1$ modes for $\tau = 0.08$ ps. The FSRS peaks reverse sign at the highest fluence of 185mJ/cm$^2$. (b) First principles calculations of the normalized Raman tensors for $Q_{IR}$ and $Q_P$ as a function of the ferroelectric polarization $P/P_0$, where $P_0$ is the value of the static polarization at room temperature and equilibrium. (c) Amplitudes of the Raman peaks associated with $Q_{IR}$ and $Q_P$. At fluences greater than ~ 60 mJ/cm$^2$, the reductions of the peak amplitudes begin to exhibit a saturation-like behavior, likely indicating a renormalization of the frequency of the driven mode due to the anharmonic coupling between $Q_{IR}$ and $Q_P$. The peak amplitudes of as a function of $\tau$ at the highest excitation fluence of 185mJ/cm$^2$ are shown for (d) $Q_{IR}$ and (e) $Q_P$. The calculations were performed in the same manner as those of Figure 4, except that here it is assumed that a transient sign reversal of the Raman tensor propagates throughout the medium, rather than a simple reduction. The results of (b) were used to model the corresponding Raman tensors for $Q_{IR}$ and $Q_P$. Details are provided in the supplementary information.

# Supplementary information for
# Nonlocal nonlinear phononics

M. Henstridge[1], M. Först[1], E. Rowe[1], M. Fechner[1], A. Cavalleri[1,2]

[1] Max Planck Institute for the Structure and Dynamics of Matter, 22761 Hamburg, Germany

[2] Department of Physics, Clarendon Laboratory, University of Oxford, Oxford OX1 3PU, United Kingdom

**S.1) Summary of the FSRS process**

FSRS requires two incident electric fields, $E_r$ and $E_w$, with respective carrier frequencies $\omega_r$ and $\omega_w$ chosen in the visible spectral range for this experiment. $E_r$ has a narrow spectral bandwidth (< 1 THz) corresponding to a pulse duration of typically 1-3 ps. $E_w$ is typically a coherent white light continuum pulse with a spectral bandwidth spanning tens of THz and temporal FWHM $\lesssim$ 100 fs. When the two pulses impinge on the LiNbO$_3$ crystal, a force $F = R^* E^2$ is then imparted onto a given mode $Q$ that has Raman activity. For the FSRS process, we are concerned with the force component proportional to $E_r E_w$ which oscillates at the difference frequency $\omega_r - \omega_w$. The relevant equation of motion for the mode $Q$ with eigenfrequency $\omega_Q$ and damping rate $\gamma_Q$ is then

$$\ddot{Q} + 2\gamma_Q \dot{Q} + \omega_Q^2 Q = R^* E_r E_w \propto \cos[(\omega_r - \omega_w)t]. \quad \text{(S1)}$$

Any mode with frequency $\omega_Q$ can be stimulated through eqn. (S1) as long as the white light continuum contains spectral weight at $\omega_r \pm \omega_Q$. The stimulated mode $Q$ then mixes with $E_r$ via the Raman tensor, generating a nonlinear polarization $P_{rQ} = \epsilon_o R^* E_r Q$ which oscillates at frequencies $\omega_r \pm \omega_Q$. The resulting scattered field $E_{sc}$ is given by the inhomogeneous wave equation

$$\nabla^2 E_{sc} - \frac{n^2}{c^2}\frac{\partial^2 E_{sc}}{\partial t^2} = \frac{1}{\epsilon_o c^2}\frac{\partial^2 P_{rQ}}{\partial t^2} = \frac{1}{c^2}\frac{\partial^2 (R^* E_r Q)}{\partial t^2}. \quad \text{(S2)}$$

Here, $n$ is the index of refraction, $c$ is the vacuum speed of light, and $\epsilon_o$ is the vacuum permittivity. Eqns. (S1) - (S2) describe a four-wave mixing process in which $E_{sc}$ is ultimately emitted in the same direction as $E_w$ [34]. When sent to a spectrometer, the two fields interfere in the frequency domain, with the resulting intensity given by:

$$I_{FSRS}(\omega) = \left|\int (E_w + E_{sc})e^{i\omega t}dt\right|^2. \quad (S3)$$

As shown in Figure 2 of the manuscript, the frequency components of $E_{sc}$ at $\omega_r - \omega_Q$ and $\omega_r + \omega_Q$ interfere with the incident field $E_w$ constructively and destructively, respectively. All FSRS spectra shown in the manuscript are normalized to the power spectrum of $E_w$ according to:

$$FSRS(\omega) = \frac{\left|\int (E_w + E_{sc})e^{i\omega t}dt\right|^2}{\left|\int E_w e^{i\omega t}dt\right|^2}. \quad (S4)$$

### S.2) Sample preparation

All measurements were performed at room temperature and in ambient air. The sample investigated was a 50μm x-cut LiNbO$_3$ single crystal which was coated on the back side with an anti-reflection layer designed for 400 nm wavelengths. This reduced the prominence of fringes in the spectrum of the white light continuum due to Fabry-Perot interferences.

### S.3) Experimental Setup

**a) mid-IR pump.**

A regeneratively amplified Ti:sapphire oscillator operating at a 1 kHz repetition rate was used for the experiments, delivering pulses of ~100 fs duration at 800 nm carrier wavelength. 4 mJ of the pulse energy was used to pump an optical parametric amplifier which produced 1.2 mJ of combined signal and idler when tuned to a signal wavelength of 1520 nm. The two near-infrared beams were then used to generate mid-infrared pulses with 12 µJ of energy and 20 THz central frequency (7 THz FWHM) by difference frequency mixing in a GaSe crystal.

**b) FSRS probe.**

To construct the FSRS probe, a portion of the remaining 800 nm output from the regenerative amplifier was split into two paths. The first path employed roughly 1µJ in order to generate a white light continuum pulse in a $CaF_2$ substrate. Using a motorized stage, the substrate was kept under constant translational motion in order to avoid damage. The spectral weight of the continuum pulse extended to 350 nm wavelengths; however, the edge pass filter used to remove the residual 800 nm light reduced the shortest usable wavelength to 388 nm. The continuum pulse was collimated and then focused onto the sample using reflective optics.

To generate the narrowband Raman pump field, 10 µJ pulses of the 800 nm was incident upon a 1.5 cm thick type-I BBO, generating 1.4 µJ of second harmonic with central wavelength of 400 nm (750 THz). The pulse envelope was characterized by cross correlation with the white light continuum pulse in a time-resolved two-photon absorption measurement in the 50 µm-thick $LiNbO_3$. The temporal width of the pulse envelope, which is shown in Figure S1, is determined by the group velocity mismatch between the fundamental 800 nm pulse and the generated 400 nm pulse in the BBO. The asymmetry in the intensity envelope results from the depletion of the 800 nm pulse as it passes through the crystal. The temporal width between the rising and falling edge of the pulse was roughly 3 ps.

**c) FSRS data acquisition**

In order to facilitate single-shot detection of the white light continuum pulses, the CCD of the visible spectrometer was triggered at the 1 kHz repetition rate of the regenerative amplifier. The narrowband

Raman pump pulses were mechanically chopped at half the frequency (500 Hz), facilitating a consecutive narrowband pulse on/off sequence impinging upon the sample. To produce a given FSRS spectrum, we collected a series of shots (minimum 50,000) and binned the data such that the sum of the "on" shots was divided by the sum of the "off" shots. The delay between the white light continuum and narrowband pulses was fixed to 1.85 ps with respect to the time trace shown in Figure S1.

When performing time-resolved FSRS measurements, the mid-IR pump illuminated the sample during both "on" and "off" shots. The sum of the "on" shots was divided by the sum of the "off" shots, in the same manner as the equilibrium FSRS measurements. Thus, effects which are dependent on the carrier-envelope phase of the mid-IR pulse, which can vary over modulo $2\pi$ from shot to shot, are averaged out in both the "on" and "off" shots. This was essential for averaging out parasitic carrier-envelope phase dependent effects of the impinging mid-IR pulses on the white light continuum spectrum.

### S.4) Consideration of the polariton and probe velocities

The velocity of the probe pulses relative to the excitations arising from the mid-IR pump was determined by considering the group refractive index of LiNbO$_3$ at 400 nm ($n_{g0} = 2.86$) [35] as well as the $\theta = 30°$ angle between the mid-IR pump and the white light continuum. In the rest frame of the polariton, the probe moves slower than it would in the case that it was collinear with the polariton; the effective group velocity is given by $v_g = c/n_g = c/(n_{g0}\cos(30°))$ where $c$ is the speed of light in vacuum. Figure S2 shows the mapping of $n_{g0}$ and $n_g$ onto the dispersion relation for the c-axis modes, which was generated by using parameters from infrared reflectivity measurements in Ref. 36. Notably, the group velocity line intersects the polariton branch of the 18.3 THz TO phonon at a frequency of 16 THz. This indicates that the probe is phase-matched to effects which result from the propagating 16 THz polariton.

The phase index at 2.5 THz is $n_p$= 5.15, yielding a ratio of $v_g/v_p$= 1.56. The corresponding velocity line was then used for the plot of Figure 4(b) and in the calculations which account for the velocity mismatch between the probe and the propagating reduction of the Raman tensor (see section S.5).

**S.5) Time-domain model of the FSRS process**

Here, we discuss the model that was used to produce the calculations shown in Figures 4(a) and 5(d)-(e) of the main text. The fields for the FSRS probe were modelled as

$$E_w(t) = E_{wo} e^{-t^2/2t_w^2} \cos(\omega_w t) \quad (S5)$$

$$E_r(t) = E_{ro} e_r(t) \cos(\omega_r t) \quad (S6)$$

$E_w(t)$ served as a model white light continuum field, with $t_w$= 3 fs to ensure a large spectral bandwidth. $E_r(t)$ modelled the narrrowband field used in the experiments, and $e_r(t)$ is an interpolation of the pulse envelope shown in Figure S1. Following the discussion presented in Section S-3, the equation of motion for the $n^{th}$ c-axis mode was modelled as

$$\ddot{Q}_n(t) + 2\gamma_n \dot{Q}_n(t) + \omega_n^2 Q_n(t) = R_n^* E_r(t) E_w(t). \quad (S7)$$

The polarization resulting from the mixing of $Q_n$ with the narrowband field via the Raman tensor is

$$P_n(t) = R_n^* Q_n(t) E_r(t). \quad (S8)$$

The radiated field $E_n(t)$ is proportional to the time-derivative of the polarization

$$E_n(t) \propto \frac{\delta}{\delta t}[R_n^* Q_n(t) E_r(t)] \quad (S9)$$

The FSRS spectrum as a function of frequency $f$ is then given by

$$FSRS(f) = \frac{|\int (E_w(t) + \sum_n E_n(t)) e^{i\omega t} dt|^2}{|\int E_w(t) e^{i\omega t} dt|^2}. \quad (S10)$$

To model the equilibrium FSRS spectrum, the parameters $R_n^*$ and $\gamma_n$ were chosen so that eq. S10 fits the experimental data.

Our model for the dynamic FSRS spectra resulting from mid-IR excitation assumes that $R_n^*$ becomes time-dependent and consists of two components:

$$R_n^*(t) = R_n^-(t) + R^{16}(t) \quad (S11)$$

Here, $R_n^-(t)$ corresponds to a transient reduction of the equilibrium Raman tensor for the n$^{th}$ c-axis mode, and $R^{16}(t)$ models the oscillatory component responsible for the second-order scattering. The shape of $R_n^-(t)$ was determined by first assuming that the ferroelectric polarization $P$ is reduced over the same time scale as the data reported in Ref. [37]. First principles calculations of the Raman tensors $R_{IR}^*(P)$ and $R_P^*(P)$ (see Figure 5(b) of the main text) were then mapped onto the time-dependent polarization in order to produce $R_{IR}^*(t)$ and $R_P^*(t)$. Our model then assumes that these Raman tensors propagate throughout a 50 µm sample volume with a group velocity equal to the soft mode polariton phase velocity. To account for the velocity mismatch between the propagating $R_n^*(t)$ and the FSRS probe, the velocity line of the probe was projected onto the space-time maps for the propagating Raman tensors, and an integral along the velocity line was carried out for each time step. This produces the effective $R_{IR}^-(t)$ and $R_P^-(t)$ which are seen in the rest frame of the probe. For each data set, the degree of polarization reduction was adjusted in order to produce a given $R_{IR}^-(t)$ and $R_P^-(t)$ which produced calculations that best fit the data. The Raman tensors of the two additional c-axis modes which appear in the Raman spectrum at 8.3 THz and 10 THz were assumed follow the same dynamics as $R_P^-(t)$.

$R^{16}(t)$ was assumed to have the form:

$$R^{16}(t) = A\left(\frac{Erf[at]}{2} + \frac{1}{2}\right) \exp[-\pi\gamma t] \cos[2\pi f t + \phi] \quad (S12)$$

where $Erf[at]$ is an error function with rise parameter $a$ that was chosen by integrating the Gaussian envelope of the mid-IR pulse used in the experiments. Additionally, $\gamma$ is the lifetime of the 18.3 THz mode extracted from infrared measurements reported in Ref. 36, $f$ was taken to be 16 THz, $A$ was chosen as a fit for each data set, and $\phi$ is the phase of the oscillation. Generally, the phase $\phi$ of a linearly driven polariton is dependent on the carrier envelope phase (CEP) of the excitation pulse. In the experiments, the CEP of the mid-IR source fluctuates randomly from shot to shot, hence the time-resolved FSRS measurements were averaged over ~ 100,000 shots. To account for the phase averaging, equations S1 – S5 were evaluated from $\phi = 0 \ to \ 2\pi$ in steps of $\pi/2$ for each delay $t_d$ between the FSRS probe and $R_n^*(t + t_d)$. The expression which describes the simulated CEP-averaged FSRS spectrum for a given $t_d$ is then:

$$I_{FSRS}(f)_{t_d} = \frac{\sum_\phi |\int (E_w(t) + \sum_n E_n(t))e^{i\omega t}dt|^2}{\sum_\phi |\int E_w(t)e^{i\omega t}dt|^2} . \quad (S13)$$

**S.6) DFT calculations of the polarization dependent Raman tensor**

The dependence of the Raman tensor elements on the polarization was mapped out by utilizing the first-principle framework of density functional theory (DFT) calculations. We performed our computations with the Vienna ab-initio simulation package VASP.6.2 [38]. Moreover, for the phonon calculation, we used the Phonopy software package [39]. Our computations further utilized pseudopotentials generated within the Projected Augmented Wave (PAW) [40] method. Specifically, we take the following configured default potentials: Li $2s^1$, Nb $4p^65s^14d^4$, and O $2s^22p^4$. We applied the PBEsol [41] approximation for the exchange-correlation potential. As a numerical setting, we used a 9x9x9 Monkhorst [42] generated k-point-mesh sampling of the Brillouin zone and a plane-wave energy cutoff of 750 eV. The self-consistent calculations were reiterated until the change in total energy becomes less than $10^{-9}$ eV. Last, to compute the Raman tensor elements, we followed the approach of Ref. [43] and determined the R* elements from the induced change in the permittivity due to phonon displacement.

The starting point of our investigation was the ferroelectric ground state of LiNbO$_3$. To create a force-free reference configuration, we first structurally relaxed the LNO unit-cell in DFT. We initialized the computation with the experimental structure with symmetry R3c of Ref. [44]. We then minimized the system until the Hellman Feynman forces on the atoms and the pressure on the unit cell became less than 0.1 eV/Ang. Tab. S1 contains the final structure. For this atomic arrangement, we then computed the phonon spectrum at the zone center and identified the four polar A$_1$ modes at the frequencies noted in the lower part of Tab S1. All results agree with experimental findings.

Next, we constructed the reversed ferroelectric (FE) polarization configuration by inverting the positions of all atoms. Further, we built the centrosymmetric state, which lies in between the two FE states. Utilizing these three states, we finally computed the atomic displacement field, which connects all three states. In Figure S3, we show the evolution of the FE polarization, calculated by the Berry-Phase Approach [45,46] along this displacement field. Note that $\lambda=0$ corresponds to the centrosymmetric state.

Finally, we applied the approach of Ref. [43] to compute the Raman tensor for the structures along this trajectory. Note that we recalculated the phonon spectrum for each structural configuration to obtain the corresponding correct eigendisplacements of the phonons. Fig. S3 (b) shows the resulting R* dependence. An apparent reduction and sign change become evident upon passing through the centrosymmetric state, with exception to the 10.1 THz mode, which remains finite through the transition. When restoring inversion symmetry, the phonon modes recapture their unique character to be either infrared or Raman active. In the centrosymmetric state, LNO exhibits three polar modes and one Raman active mode, which become the c-axis polar modes in the ferroelectric phase. The 10.1 THz mode corresponds to the original Raman mode, which maintains its Raman activity, and thus a nonzero R*, in the centrosymmetric state.

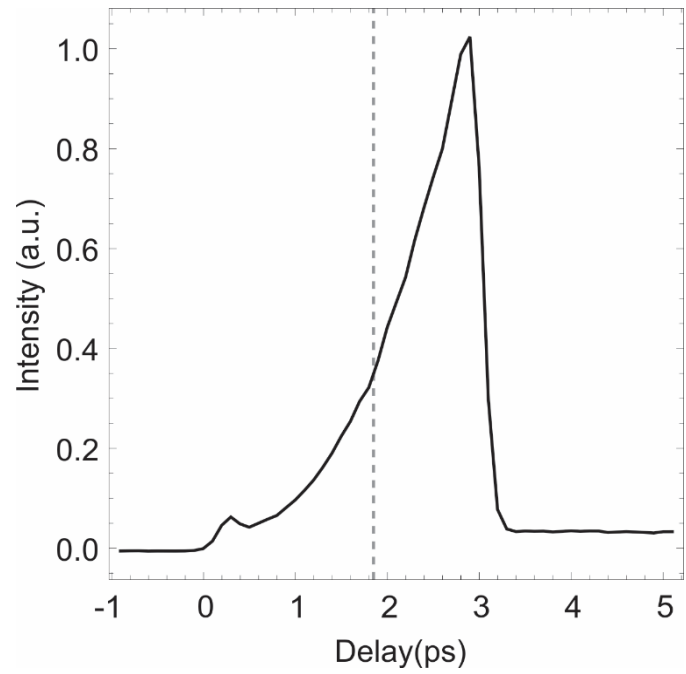

**Figure S1.** Intensity envelope of the narrowband field used in the FSRS probe, as obtained from two-photon absorption measurements in the 50 µm thick $LiNbO_3$ sample. The dashed line indicates the fixed position of the white light continuum pulse during the FSRS measurements.

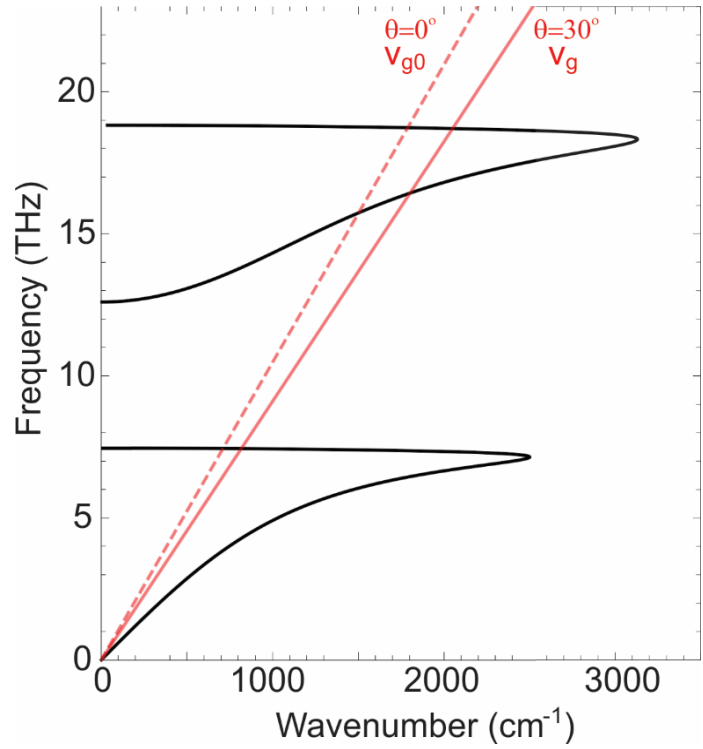

**Figure S2.** Black curve: Dispersion relation for the c-axis modes in LiNbO$_3$ plot using parameters obtained from Ref. 36. The slope of the dashed red curve indicates the group velocity of a pulse with carrier wavelength of 400 nm, and the solid red curve indicates the effective group velocity of such a pulse propagating at an angle of 30° relative to the polariton.

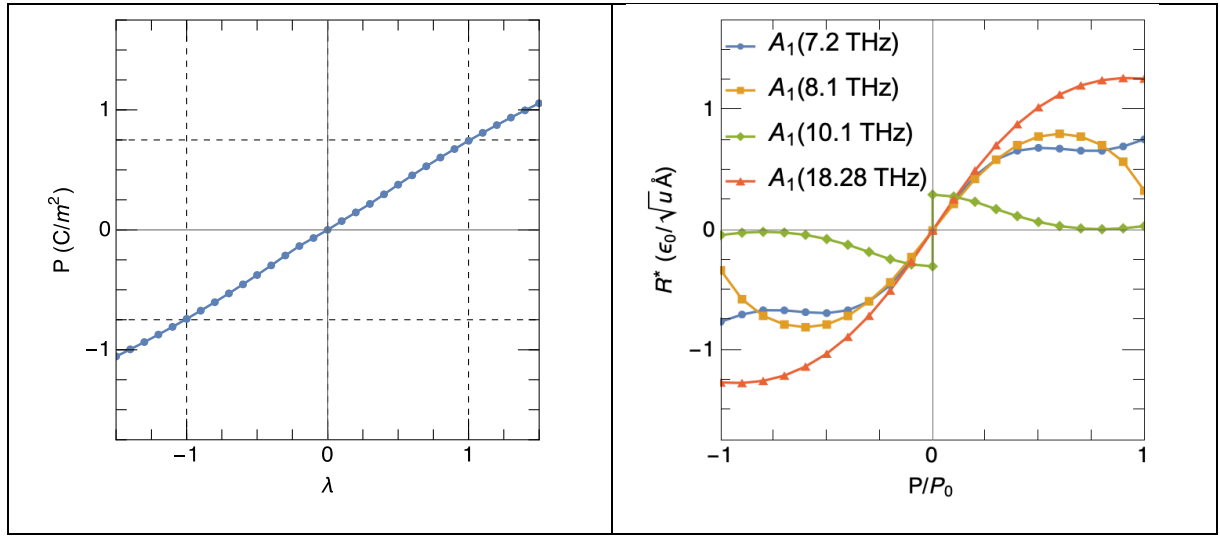

**Figure S3.** Left Panel: Total polarization as a function of structural distortion going from the $-P_0$ ferroelectric state to the inverted $+P_0$ state crossing the centrosymmetric state in the middle. Right panel: Raman tensor of the four polar $A_1$ modes in $LiNbO_3$

| R3c | Theo. | Expt.[44] | | | R-3C | | | |
|---|---|---|---|---|---|---|---|---|
| $a_{hex}$ | 5.13 Å | 5.15 Å | | | 5.13 Å | | | |
| $c_{hex}$ | 13.81 Å | 13.86 Å | | | 13.81 Å | | | |
| Atomic positions: | | | | | | | | |
| | Wykoff pos. | x | y | Z | Wykoff pos. | x | y | Z |
| Li | 6a | 0 | 0 | 0.78902 | 6a | 0 | 0 | 0.25 |
| Nb | 6a | 0 | 0 | 0.00674 | 6b | 0 | 0 | 0 |
| O | 18b | 0.04711 | 0.34397 | 0.07142 | 18e | 0.61996 | 0 | 0.25 |
| Phonons: | | | | | | | | |
| phonon mode | Theory [THz] | Exp.[47] [THz] | | | | | | |
| A1 | 7.20 | 7.55 | | | | | | |
| A1 | 8.10 | 8.24 | | | | | | |
| A1 | 10.06 | 9.95 | | | | | | |
| A1 | 18.28 | 18.95 | | | | | | |

**Table S1.** Structural parameters in the hexagonal setting and phonon frequencies from DFT (this work) and experimental results.

**References for the Supplementary Information**